\documentclass{article}
\usepackage{epsfig}
\usepackage{amssymb}
\usepackage{amsmath}
\usepackage{amsfonts}
\newcommand \be{\begin{eqnarray}}
\newcommand \ee{\end{eqnarray}}
\begin{document}
\begin{center}
{\bf Spin 1/2 Fermions in the Unitary Limit.I}\\
\bigskip
\bigskip
H. S. K\"ohler \footnote{e-mail: kohler@physics.arizona.edu} \\
{\em Physics Department, University of Arizona, Tucson, Arizona
85721,USA}\\
\end{center}
\date{\today}

\begin{abstract}
This report concerns the energy of a zero-temperature many-body system of 
spin $\frac{1}{2}$ fermions interacting via a two-body potential 
with a free space infinite scattering length and zero effective range; the
Unitary limit.
Given the corresponding phase-shift $\delta(k)=\pi/2$ a one-term separable
potential is obtained by inverse scattering assuming 
a momentum cut-off  $\Lambda$ such that $\delta(k)=0$ for $k>\Lambda$.
The \it effective \rm interaction in the many-body system is calculated 
in a pp-ladder approximation with Pauli-blocking but
neglecting mean-field (dispersion) corrections; effective mass
$m^{*}=1$.
Using only the zero relative momentum component of this interaction
the total energy is $\xi=4/9$ (in units of the fermigas), a result
reported by several previous authors.  Integrating the momentum dependent
interaction over the Fermi sea this energy is revised to $\xi=0.24.$
This result is independent of density and of the cut-off $\Lambda$ if
$\Lambda > \sim 3k_{f}$.

With $m^{*}\neq 1$ there is however a strong dependence on this cut-off.

Including hh-ladders estimates give $\xi=0.4\leftrightarrow 0.6$, but
a reliable result would in this case require a Green's function calculation.

\end{abstract}

\section{Introduction}
The properties of  a dilute fermigas with large scattering length 
is of considerable theoretical as well as experimental interest. 
Taking advantage of Feshbach resonances it is  possible to
magnetically tune the atomic scattering lengths,
e.g.\cite{reg05,geh03,bar04}.
Increasing the scattering length of fermions
from $-$ to $+\infty$ resulting in bound boson systems
to explore the crossover from BCS to BEC has been reported  by several
groups. 

 A theoretically related  problem proposed
by George Bertsch \cite{ber99} is that of the energy of a 
dilute system of spin 1/2 fermions
interacting via a zero-range, infinite scattering length
interaction referred to as the Unitary limit.
For such a system one would expect the existence of a constant $\xi$ 
being a function only of fundamental constants such that the total 
energy $E=\xi E_{FG}$ where $E_{FG}$ is the uncorrelated Fermi-gas
energy.

Several numerical methods have been used to determine $\xi$.
The Monte Carlo calculations of
Carlson et al.\cite{car03} are generally regarded to be the most complete
and gives $\xi=0.44 \pm 0.01$. The same (or similar)
results are showm in refs. \cite{per04,ste00,che06,mos06} some of which
are $\xi=4/9$.
Other authors report values of $\xi=0.326$ \cite{hei01,bak99} while a
recent result is $\xi=0.25$\cite{lee06}.

It is a well-known fact that an interaction with large scattering length is
separable.
This paper is a  report on results of  calculations
using a one term separable
two-body interaction determined by inverse scattering from the phase-shift
$\pi/2$, i.e. with infinite scattering length and zero effective range.
It is shown in Section 2 that the calculations are greatly
simplified in the Unitary limit with an effective mass $m^{*}=1$.

Numerical results   are shown in Section 3. 
In the limit when the
Pauli blocking $Q\rightarrow 1$, the theory reduces to the phase-shift
approximation as shown in Section 4.
A comparison with some of the results of other authors is shown in Section
5 and a short summary with comments is found in Section 6.

\section{Separable Interaction;Formalism}
The use of a separable interaction in Nuclear Physics problems has a long
history. It seems however that the first consistent calculation using
inverse scattering techniques to construct a separable NN potential 
with an application to the nuclear matter
problem was that reported in ref. \cite{kwo95}. A close agreement
with calculations using the meson-theoretical potentials of Machleidt was
found.  
Subsequent use was shown
in ref. \cite{hsk04,hsk06} relating to $V_{low-k}$ etc. In the latter of
these two last references the dispersion corrections and its relation to
saturation was of primary interest.
It is well known that for a two-particle system with a bound state at or
close to zero energy the interaction can be represented by a separable
potential.
The method described below in which this separable potential
is obtained by inverse scattering is therefore suitable when
considering the problem at hand, large scattering lengths. 
One of the main problems in an inverse scattering calculation 
is the change in sign of the phasehift as a
function of relative momentum.  In the present case 
this is not an issue.

A rank 1 separable potential provides a sufficient and in fact precise
description of the interaction in the Unitary limit. 
It is assumed to be attractive and given by

\begin{equation}
V(k,p)=-v(k)v(p)
\label{V}
\end{equation}
Inverse scattering then yields
(e.g. ref \cite{kwo95,tab69})

\begin{equation}
v^{2}(k)= \frac{(4\pi)^{2}}{k}sin \delta (k)|D(k^{2})|
\label{v2}
\end{equation}
where
\begin{equation}
D(k^{2})=exp\left[\frac{2}{\pi}{\cal P}\int_{0}^{\Lambda}
\frac{k'\delta(k')}{k^{2}-k'^{2}}dk' \right]
\label{D}
\end{equation}
where ${\cal P}$ denotes the principal value  
$\delta(k)$  the phaseshift. $\Lambda$ provides a cut-off in
momentum-space. 
The effect of the  cut-off will be exploited below.

With $\delta(k)=\pi/2$ one finds 
\begin{equation}
v^{2}(k)= -\frac{(4\pi)^{2}}{(\Lambda^{2}-k^{2})^{\frac{1}{2}}}
\label{vpi2}
\end{equation}
and the interaction reduces to a constant for $\Lambda \gg k$, but
$\rightarrow -\infty$ for $k\rightarrow \Lambda$.

The diagonal elements of the in-medium
 interaction is

\begin{equation}
 G(k,P)=-\frac{v^{2}(k)}{1+ I_{G}(k,P)}
\label{G}
\end{equation}
with
\begin{equation}
I_{G}(k,P)=\frac{1}{(2\pi)^{3}}\int_{0}^{\Lambda}
v^{2}(k')\frac{m^{*}Q(k',P)}{k^{2}-k'^{2}}
k'^{2}dk'
\label{I_G}
\end{equation}
where $P$ is the center of mass momentum,  $Q$ the angle-averaged
Pauli-operator for pp-ladders (Brueckner approximation) 
and $m^{*}$ is the effective mass. One should note that 
the angle-averaging  is exact in the effective mass approximation. 

The divergence of
$v^2(k)$ for large $k$ indicated after eq.(\ref{vpi2}) makes the 
numerical integration in eq.(\ref{I_G}) somewhat complicated. This can be
overcome by the substitution $k'=\Lambda sin(t)$ to get
\begin{equation}
I_{G}(k,P)=\frac{1}{(2\pi)^{3}}\int_{0}^{\pi/2}
\frac{m^{*}Q(\Lambda sin(t),P)sin^{2}tdt}{(k/\Lambda)^{2}-sin^{2}t}
\label{I_Gt}
\end{equation}

The effective interaction $G(k,P)$ is in principle $\Lambda$-dependent.
With $m^{*}=1$ and $k/\Lambda\rightarrow 0$ and $Q\rightarrow 1$ 
for $\Lambda\gg k_{f}$, $I_{G}(k,P)\rightarrow -1+{\cal O}(1/\Lambda)$. 
With $v^{2}(k)\rightarrow 1/\Lambda$ one therefore finds $G(k,P)$ independent of
$\Lambda$ for large $\Lambda$.
If on the other hand one sets $1/\Lambda=0$ in eq. (\ref{I_Gt}) 
then $G\equiv K$,  where
$K$ is the reactance matrix as defined below. 

The (in)dependence of $\Lambda$ is more clearly seen by using the
following method  applicable with $m^{*}=1$. One should note that both of
the eqs (\ref{I_G}) and (\ref{I_Gt}) involve integrations up to $\Lambda$.
A regularization that restricts the
momentum-integration in eq. (\ref{I_G}) to momenta $\leq 2k_{f}$  can 
be achieved as follows.

With a separable interaction the Reactance-matrix $K$ is defined by
\begin{equation}
 K(k)=-\frac{ v^{2}(k)}{1+ I_{K}(k)}
\label{K}
\end{equation}
with
\begin{equation}
I_{K}(k)=\frac{1}{(2\pi)^{3}}\int v^{2}(k')\frac{\cal P}{k^{2}-k'^{2}}
k'^{2}dk'
\label{I_K}
\end{equation}
where $\cal{P}$ refers to a principal value integration.
Therefore $K$ is real and its on-shell diagonal component is given by
\begin{equation}
K(k)=-4\pi tan\delta(k)/k.
\label{Kk}
\end{equation}
 
In the Unitary case with $\delta(k)=\pi/2$, $K(k)\rightarrow \infty$,
implying $I_{K}(k)=-1$. As stated above this is consistent 
with eq. (\ref{I_Gt}) which for
$Q=1$ and $m^{*}=1$ also results in $I_{G}=-1$ independent of $k<\Lambda$.

Eqs (\ref{G}) and (\ref{K}) can be combined to get \footnote{This
is a somewhat similar subtraction method as used to get eq. V(34) 
in ref.(\cite{gebrown}) and also used by S.A. Moszkowski in
unpublished work \cite{mos06}.}
\begin{equation}
 G(k,P)=-\frac{v^{2}(k)}{I_{GK}(k,P)}
\label{GK}
\end{equation}
with
\begin{equation}
I_{GK}(k,P)=\frac{1}{(2\pi)^{3}}\int
v^{2}(k')\frac{Q(k',P)-{\cal P}}{k^{2}-k'^{2}+i\eta}
k'^{2}dk'+\frac{kv^{2}(k)}{tan\delta(k)}
\label{I_GK}
\end{equation}
The resulting form, eq.(\ref{GK}) is very suitable  for the problem at hand because
in the unitary limit the last term in eq. (\ref{I_GK}) $\rightarrow 0$.
Furthermore, $Q\rightarrow 1$ for $k'>2k_{f}$. 
The summation (integration) over intermediate states in eq. (\ref{I_GK})
therefore only involves  momenta $\leq 2k_{f}$. 
One should observe that with $Q$ defined for pp-ladders, $I_{GK}$ is real 
for $k<k_{f}$ which is the case here. With hh-ladders one has pole-terms
contributing to imaginary parts.

Note that this regularization of the $G$-matrix equation 
is possible only because of the neglect of the dispersion 
correction i.e. with $m^{*}=1$.

As shown by eq. (\ref{vpi2}), $v(k)$ is constant to any desired accuracy
for momenta $k\leq 2k_{f}$ by choosing $\Lambda$ large enough. 
Equation (\ref{GK}) then simplifies. It is independent of the interaction
because one can set $v(k)=v(k')$. The momentum-integration can  then be 
done analytically as shown below.
This simplification is not possible if using 
eq. (\ref{I_G}) with integration up to the cut-off $\Lambda$ where the
interaction (\ref{vpi2}) diverges. It is also simpler than although in principle
equivalent to eq.  (\ref{I_Gt}).

As stated above, the Reactance matrix is real 
and so is the $G$-matrix (for occupied states) when
defined with pp-ladders like in Brueckner theory.
After dividing by $v^{2}(k')=v^{2}{k}$ in  eq. (\ref{I_GK}) the integral 
is conveniently evaluated analytically 
One finds 
with $a=\frac{k}{k_{f}}$ and $y=\frac{P}{2k_{f}}$:
\begin{equation}
I_{GK}(a,y)=\frac{k_{f}}{\pi}\left[1+y+a*log\left|\frac{1+y-a}{1+y+a}\right|
+\frac{1}{2y}(1-y^{2}-a^{2})log\left|\frac{(1+y)^{2}-a^{2}}{1-y^{2}-a^{2}}
\right|\right]+kcot\delta(k).
\label{Ipp}
\end{equation}

 To include hh-ladders  there is an additional term
\begin{equation}
I_{hh}(a,y)=\frac{k_{f}}{\pi}\left[2(1-y^{2})^{\frac{1}{2}}+a*
log\left|\frac{(1-y^{2})^{\frac{1}{2}}-a}{(1-y^{2})^{\frac{1}{2}}+a}\right|\right].
\label{Ihh}
\end{equation}

The diagonal $G$-matrix elements are
\begin{equation}
G(a,y)=-4\pi[I_{GK}(a,y)-\frac{1}{a_{s}}+\frac{1}{2}r_{0}k^{2}]^{-1}
\label{Gpp}
\end{equation}
where the two terms with scattering length $a_{s}$ and effective range $r_{0}$ 
drop  out in the unitary limit but kept here for a
discussion in Sect (5). 

With $G(a,y)$ given, the potential energy per particle $PE/A$ is 
\begin{equation}
PE/A=\frac{3k_{f}^{3}}{\pi^{2}}\int_{0}^{1}\left[\int_{0}^{1-a}8G(a,y)y^{2}dy+
\frac{1}{a}\int_{1-a}^{(1-a^{2})^{\frac{1}{2}}}4G(a,y)(1-y^{2}-a^{2})ydy\right]a^{2}da
\label{PE}
\end{equation}

The kinetic energy per particle, i.e. the uncorrelated fermi-gas energy 
is given by $$E_{FG}/A=\frac{3}{10}\frac{\hbar^{2}}{m}k_{f}^{2}.$$
The total energy is expressed in these units by
$$E/A=\xi E_{FG}/A.$$

\section{Numerical Results}
All results are for allowing the scattering length $a_{s}\rightarrow
\infty$ and effective range $r_{0}=0$.
The integral $I(a,y)=0$ (i.e. $G\rightarrow \infty$) along a line $y=f(a)$ in the $(a,y)$-plane from
approximately $(0.84,0)$ to approximately $(0.90,0.30)$. This complicates
the numerical evaluation of the potential energy. This line is first extracted
numerically by iteration at each meshpoint of the variable $a$ to find
$f(a)$. 
The function $I(a,y)$ is fitted to second order in $y$ for each value
of $a$ in some  
interval $\Delta y$ across this line $f(a)$. The same techniques as used in
standard principal value integrations is then used together 
with a shift in meshes so that one point will  be located on the  line $f(a)$.
 
The momentum-integration leading to the effective
interaction was calculated analytically using eq. (\ref{Ipp}) and 
also numerically from eq. (\ref{I_Gt}) (with the same result).
The integrations for $PE/A$, eq. (\ref{PE}), were done numerically.
The result  independent of density is

 $$\xi=0.24.$$

 



It may be of some interest to see some results of approximations. 
So for example with $$I(a,y)\rightarrow
I(0,0)=-\frac{2k_{f}}{4\pi^{2}}$$ one finds
$$\xi=4/9$$  which is a result also obtained by Steele
\cite{ste00} in the same approximation, but quite different from our
$\xi=0.24$.
If on the other
hand one allows for a dependence on center of mass momentum by using
$I(o,y)$ one finds
$$\xi=0.515.$$
One has to conclude that the momentum dependence on the effective interaction
$G(k,P)$ 
cannot be ignored when calculating the energy of the many-body system.
 
The above results are were obtained with a summation over pp-ladders as in
Brueckner theory. One may inquiry as to the importance of hh-ladders 
for the present problem. By redefining the $Q$-operator by using $$Q=1-n_{1}
-n_{2}$$ instead of $Q=(1-n_{1})(1-n_{2})$ ($n$ being occupation-numbers)
in the above equations one finds
 $$\xi=0.4 \leftrightarrow 0.6.$$
The uncertainty in the results is related to the pairing instability.
Experience from nuclear matter calculations
appears to be that this is to a large extent resolved by the Green's
function method with integrations over the energy variable in the spectral
functions.
The above estimate of including the hh-ladders cannot be considered
meaningful.
A Green's function calculation is necessary in this case; results
will be presented in a forthcoming report.

To some approximation one might expect the pp- and hh-ladder contributions
to be equal. This assumption was used in ref.\cite{hei01}. Doing so here
one finds $$\xi=0.56.$$
But again, this result is not reliable.
 
 The convergence of $G(k,P)$ when increasing $\Lambda$ is not obvious
from eqs (\ref{G}-\ref{I_Gt}).
These eqs are however equivalent 
with eqs (\ref{GK},\ref{I_GK}) which clearly do converge. That these two
set of equations are indeed equivalent was also found numerically.
But the last set of equations assume that $m^{*}=1$. It is found from
numerical tests that eqs (\ref{G}-\ref{I_Gt}) do not converge with
increasing $\Lambda$ so the problem with $m^{*}\neq 1$ is not resolved here. 
As an example it is found that with $m^{*}=0.9$, $\xi\rightarrow 1$ 
as $\Lambda \rightarrow 10^{3}k_{f}$.

\section{$Q\rightarrow 1$; The Phase-shift approximation}
 Two particles in a large box can be considered a limiting case of a
many-body system. The effective interaction between two particles in a box
having a relative momentum $k$ is  known to be given by
\cite{dew56,fuk56,ries56,uhl36,got66}
$$G(k)=-4\pi\delta(k)/k.$$ 
It is reasonable to expect a proper many-body theory of effective
interactions to give this result in the limit $Q\rightarrow 1$. 
This limit has to be taken with caution. If one simply
lets $Q \rightarrow 1$ in eq. (\ref{I_GK}) and then takes the principal value
one finds $$I_{GK}=\frac{kv^{2}(k)}{tan\delta(k)}$$
and 
$$G_{Q\rightarrow 1}(k)=-4\pi tan\delta(k)/k\equiv K(k)$$ 
where $K(k)$ was defined by eq. (\ref{Kk}) that relates to the \it free \rm
scattering of two particles rather than two particles in a box. (If instead
inserting an $i\eta$ in the denominator one obtains the complex
$T$-matrix.)
The correct limit is obtained after
realising that when studying a many-body system one has
to explicitly consider an enclosure of the particles in a large but finite box
with a disrete rather than continuous spectrum. 
When extending the system so that sums can be replaced by integrations 
the transition from the discrete problem to the continuous has to be  done
with care.
It has been shown \cite{dew56,fuk56,ries56} that when taking 
the limit $Q\rightarrow 1$ one should in this case use
$$\frac{1}{e}\rightarrow \frac{\cal P}{e}+\gamma$$ with
$$\gamma=k(\frac{1}{\delta(k)}-\frac{1}{tan\delta(k)}).$$
Doing so in eq. (\ref{I_GK}) one finds correctly
$$G_{Q\rightarrow 1}(k)=-4\pi \delta(k)/k.$$
This may serve as a test of the equations. 
It was used in early
work as an approximation to $G(k)$ and referred to as the \it phase-shift
approximation \rm. It was believed to be a good approximation
at low densities, substantiated by some numerical results. 
Of particular interest in relation to the Unitary problem
is that it was used to calculate the binding energy for a neutron-gas at low
density \cite{bru60,mos60}. In this approximation of the effective two-body
in-medium interaction one finds with $\delta(k)=\frac{\pi}{2}$,  $PE/A=-4/3
E_{FG}$ giving
$$\xi=-\frac{1}{3}$$ quite different from any
other result suggesting it to be a poor approximation here. The effect of the
Pauli-blocking was however also calculated in ref.\cite{mos60}. (The total
energy of the neutron-gas reported there is some $20\%$ lower 
than reported in recent
publications\cite{sch05}.) Contrary to other beliefs the
potential energy was found to be over-estimated by the phase-shift
approximation by a factor varying between $1.75-1.95$ for 
$0.1 fm^{-1}\geq k_{f} \leq 0.5 fm^{-1}$. These results were obtained with
$^{1}S_{0}$ phase-shifts and would not be directly applicable to the
problem at hand. Using the same correction for the present Unitary
problem would however give $PE/A=(-0.76\leftrightarrow -0.68)E_{FG}$ and
$$\xi=0.24\leftrightarrow 0.32$$ in fair agreement with our result
$\xi=0.24$. 

\section{Comparison with previous work}
The first publication relevant for comparing with the present work
appears to be that of Baker\cite{bak99}. He considered an attractive
square-well potential with a radius $c\rightarrow 0$ and an
extrapolation of scattering length $a\rightarrow -\infty$. 
The energy of
the system was calculated in a pp-ladder approximation similar to the
Brueckner $G$-matrix as used in the present report. It is however
modified to avoid the Emery singularities \cite{eme58}. The numerical
evaluation of the energy using this 'R'-matrix 
gives  a divergent result for
$k_{F}c\rightarrow 0$ present at all scattering lengths. A Pad\'{e}
approximant gave $\xi \sim 0.40$. Baker also provides a series
expansion of the ladder sum for $c=0$. A [2/2] Pad\'{e} approximant of
this sum gives $\xi=0.568$ while the [1/1] gives $\xi=0.326$. 

Heiselberg \cite{hei01} has made extensive calculations both using a
Galitskii resummation of hh- and pp-ladders resulting in $\xi=0.33$ and in
a low order variational caculation resulting in $\xi=0.46$.
The Galitskii method is somewhat similar to the present. One difference is
that an average momentum was used when calculating the energy.
As already mentioned above the hh- and pp-contribution were assumed to be
equal which is another approximation.

The Monte Carlo calculations of Carlson et al\cite{car03} find a large
pairing gap and a $\xi=0.44\pm 0.01$ including the pairing
contributions. Without a pairing trial function (using a Slater
determinant) they obtain $\xi=0.54$. \footnote{I am indebted to K.E.
Schmidt to point this out to me.} 

Eq. (\ref{Ipp}) for $I_{GK}(a,y)$ is the same function as $f(\kappa,s)$
(except for the $cot$-term)
in the report by Steele, although with at least
formally a very different method, using Effective Field Theory and Power
Counting.  \cite{ste00}
The expression for the potential energy (\ref{PE})
is consequently (in the limit of large sacttering length and with
pp-ladders) also equal to that of Steele's. 
In his eq. (27) he lets $f(\kappa,s)\rightarrow 2$ which
is equivalent to our $I(0,0)=\frac{2k_{f}}{\pi}$ already considered as an
approximation in Sect. 3. In this limit our
results conequently agree giving $\xi=4/9$. This result is also obtained by
Moszkowski\cite{mos06}. 
Within the framework of Steele's work his approximation
$f(\kappa,s)\rightarrow 2$ seems formally consistent with his expansion
to order $1/{\cal D}$ .
In the present work there is however no expansion
other than in $\Lambda$ and $a_{s}$ and the
integrations over $(a,y)$ gives a substantial correction from $\xi=4/9$ to
$\xi=0.24$.

Chen \cite{che06} like several other autors, finds $\xi=4/9$ but only 
in a low density expansion. He uses a relativstic approach motivated by the
analogy with the infrared limit of Coulomb correlations. In the
non-relativstic limit he  finds  the energy per particle to be a  function of
$k_{f}$ . 

Our result with pp-ladders, $\xi=0.24$ is appreciably
smaller than in most reports. 
As mentioned in Sect. 4 it is to some extent substantiated by comparison
with the neutron-gas calculations in ref. \cite{mos60}.
The calculations of Carlson et al\cite{car03}
suggest that the pairing correction would give an even smaller $\xi$. 
The theoretical value closest to our result appears to be that of
Lee\cite{lee06}  who
calculated $\xi$ on the lattice with 22 particles in a periodic cube and
found $\xi=0.25(3)$. 
The estimate alluded to in Sect. 3 including the hh-ladders  giving
$$\xi=0.4 \leftrightarrow 0.6$$
lies in the range of most of the results but needs a better treatment by
Green's function methods.

There are experimental results reported between $\xi=0.74\pm0.07$\cite{geh03} 
and $\xi=0.32^{+0.13}_{-0.10}$\cite{bar04}.

\section{Summary and discussion}
This work  addresses the problem of the energy of a zero
temperature fermion gas in the unitary limit.
The free-space two-body interaction is assumed to be separable. 
This assumption should in itself not
affect the results that are expected to be independent of the shape or
strength of the interaction defined only by the scattering length and
effective range. The details of the interaction does not otherwise enter
explicitly
into the expression for the effective interaction or many-body energy.  
The separable interaction was in eqs. (1-4) determined by inverse
scattering. Eq. (\ref{vpi2}) showed explicitly that in the unitary limit, 
$a_{s}\rightarrow \infty$ and $r_{0}\rightarrow 0$ the interaction is constant,
independent of momenta for  $k\ll
\Lambda$. As $\Lambda$ can be chosen arbitrarly large the interaction can
then, with any chosen accuracy be considered a constant for all $k\leq
2k_{f}$, $2k_{f}$ being the maximum relative momentum in  $Q-1$ in eq.
(\ref{I_GK}).
To perform the calculation of $\xi$ in the Unitary limit the effective
interaction $G$ can therefore be \it chosen \rm  to be 
independent of the interaction.

The terms with $a_{s}$ and $r_{0}$ 
in eq. (\ref{Gpp}) do depend on the interaction. This equation would  only
be approximate for finite values of these quantities and 
justified only to the degree that the interaction $v(k)$ is constant. 
Numerical solution of eqs (1-3) shows that a necessary condition is that, 
as expected,  $r_{0}=0$.

The calculations are mainly analytical. The numerical
integrations are simple. The result is as expected for a unitary limit
independent of the assumed shape or strength of the interaction as well as
of the density. The assumption of infinite scattering length  sets the
scale.

The expression in eq. ({\ref{PE}) for the potential energy agrees 
with that of Steele\cite{ste00} using an EFT
method and powercounting. This is not circumstantial, but ceratinly rooted
in the fact that both methods, the separable potential and the
EFT-power-counting rely on the nearly bound state with a pole near the real
axis. There is a difference in final
result in that the calculation here is carried a step further by doing
the momentum integrations over occupied states.

A selfconsistent inclusion of the hh-
ladders and the related  effect of spectral widths would be achieved in a Green's
function calculation. In nuclear matter calculations these effects are
found to be repulsive relative to the Brueckner results that include only
pp-ladders with the potential energy increased by $\sim 8 \% $.\cite{fri03}
Applying such a correction here results in $\xi \sim 0.3.$

Green's function methods will be used in a forth-coming report on spectral
functions and densities in momentum space.

There does not seem to be a consensus within the different theoretical
calculations. It would be expected that the most accurate are those of
Carlson et al \cite{car03}. The result of the present investigation does
not support those findings.

The results within the present formalism are independent of density. This
is expected to be general in the Unitary limit.
At higher densities traditional many body theories predict
higher order diagrams such as propagator modification by
the mean field, included in standard Brueckner calculations to become
important and. It is not clear how to deal with these problems in the
Unitary limit.

An example of such a problem is the effective mass. All calculations here are
with $m^{*}=1$. The formalism used in the present calculations only works
for this case. It is found however that the mean field is practically
independent of momentum so that $m^{*}=1$ is realistic.

BCS-pairing has not been included here. It is expected to lead to important
corrections.  Superfluid gaps have been
calculated \cite{hei01}  and are large for $k_{F}a > 1$. QMC
calculations show  $\xi$ to decrease by about $0.1$ when
BCS-correlations are included in the trial wave-function \cite{car03}.

The rather different results obtained for the coefficient $\xi$ in the
Unitary limit both in experimental as well as theoretical reports suggests
that this problem is still not resolved.

It is somewhat intriguing that although $\xi$ is expected
to be a universal constant its theoretical determination 
requires relatively complicated calculations.
It is true that some estimates such as $\xi=4/9$ 
are the result of 
very simple assumptions and agree closely with the supposedly most accurate
determination, ref \cite{car03}. Closer examination seem to suggest
however (as in the
present investigation) that these simple assumptions are not valid.

\vspace{2cm}
It is a pleasure to thank Prof. Nai Kwong for many helpful discussions and Prof.
Steve Moszkowski and Henning Heiselberg for some helpful suggestions and
information.


\begin{thebibliography}{10}
\bibitem{reg05} C.A. Regal, M. Greiner, S. Giorgini, M. Holland and D.S.
                Jin,
		Phys. Rev. Lett. {\bf 95} (2005) 250404;
                M. Greiner, C.A. Regal and D.S. Jin, 
                Nature (London) {\bf 426} (2003) 537;
		M.W Zwierlein, C.A. Stan, C.H. Schunck, S.M.F. Raupach, S.
		Gupta, Z. Hadzibabic and W. Ketterle,
		Phys. Rev. Lett. {\bf 91} (2003) 250401.
		T. Bourdel, L. Khaykovich, J. Cubizolles, J. Zhang, F.
		Chevy, M. Teichmann, L. Tarruell, S.J.J.M.F. Kokkelmans and
		C. Salomon,
                Phys. Rev. Lett. {\bf 93} (2004) 050401.
\bibitem{geh03} M.E. Gehm, S.L. Hemmer, S.R. Granade, K.M. O'Hara and J.E.
                Thomas,
		Phys. Rev. A {\bf 68} (2003) 011401(R).
\bibitem{bar04} M. Bartenstein, A. Altmeyer, S. Riedl, S.Jochim, C. Chin,
                J. Hecker Denschlag and R. Grimm,
	        Phys. Rev. Lett. {\bf 92} (2004) 120401.
\bibitem{ber99} G. F. Bertsch,
                http://www.phts.washington.edu/~mbx/george.html.
\bibitem{car03} J. Carlsson, S.-Y. Chang, V.R. Pandharipande, and K.E.
                Schmidt, 
                Phys. Rev. Lett. {\bf 91} (2003) 050401.
\bibitem{per04}	A. Perali, P. Pieri and G.C. Strinati,
                Phys. Rev. Lett. {\bf 93} (2004) 100404.
\bibitem{ste00} James V. Steele,
                nucl-th/0010066.
\bibitem{che06} Ji-sheng Chen,
                nucl-th/0602065.
\bibitem{mos06} S.A Moszkowski, Private Communication.
\bibitem{hei01} H. Heiselberg,
                Phys.Rev. A {\bf 63} (2001)043606;
		cond-mat/0307726.
\bibitem{bak99} G.A. Baker,
                Phys.Rev. C {\bf 63} (1999) 054311.
\bibitem{lee06} James Lee,
                Phys Rev. B {\bf 73} )2006) 115112.
\bibitem{kwo95} N.H. Kwong and H.S. K\"ohler, 
                 Phys. \ Rev. C {\bf 55} (1997) 1650.
\bibitem{hsk04}  H.S. K\"ohler ,
                nucl-th/0511030.
\bibitem{hsk06}  H.S. K\"ohler and S.A. Moszkowki,
                nucl-th/0703093.
\bibitem{tab69}  Frank Tabakin,
                Phys. Rev. {\bf 177} (1969) 1443.
\bibitem{gebrown} G.E. Brown and A.D. Jackson, "The Nucleon-Nucleon
                  Interaction", North-Holland 1976.
\bibitem{dew56} Bryce S. DeWitt,
                Phys. Rev. {\bf 103} (1956) 1565.
\bibitem{fuk56} N. Fukuda and R.G. Newton,
                Phys. Rev. {\bf 103} (1956) 1558.
\bibitem{ries56} W.B. Riesenfeld and K.M. Watson,
                Phys. Rev. {\bf 104} (1956) 492.
\bibitem{uhl36} E.Beth and G.E. Uhlenbeck,
                Physica  {\bf 3} (1936) 729;
                {\bf 4} (1937) 915.
\bibitem{got66} Kurt Gottfried,
                Quantum Mechanics (W.A.Bejamin, Inc.,1966) p.381.
\bibitem{bru60} K.A. Brueckner,John L. Gammel and Joseph T. Kubis,
                Phys. Rev. {\bf 118} (1960) 1095.
\bibitem{mos60} P.C. Sood and S.A. Moszkowski,
                Nucl. Phys. {\bf  21} (1960) 582.
\bibitem{sch05} A. Schwenk and C.J. Pethick,
                Phys. Rev. Lett. {\bf 95} (2005) 160401.
\bibitem{eme58} V.J. Emery, Nucl., Phys. {\bf 12} (1958) 69.
\bibitem{fri03} T. Frick and H. M\"uther,
                Phys.Rev. C {\bf 68} (2003) 034310.

\end{thebibliography}
\end{document}